\documentclass{rspublic}
\usepackage{amsmath}
\usepackage{amsfonts}
\usepackage{amssymb}
\usepackage[Symbol]{upgreek}
\usepackage[nointegrals]{wasysym}
\usepackage{mathrsfs}
\usepackage{graphicx}
\newcommand{\real}{{\mathbb R}}
\begin{document}

\title[Magnetohydrodynamics equations in $\real^3$]{\large On the Well-posedness of Magnetohydrodynamics Equations for Incompressible Electrically-Conducting Fluids}

\author[F. Lam]{F. Lam}

\label{firstpage}

\maketitle

\begin{abstract}{Navier Stokes Equations; Vorticity Equation; Maxwell Equations; Magnetohydrodynamics; Turbulence; Viscosity; Randomness}

The present paper deals with the Cauchy problem of the equations in magnetohydrodynamics in the whole space. We are concerned ourselves with incompressible, electrically-conducting fluids. 
It is shown that the equations of motion are globally well-posed for any initial smooth and localized data. The mathematical structure of the solution shows that the combined magneto-fluid has the characters of turbulence, if the initial data exceed certain size. In addition, consideration has been given to a number of cases where good approximations to the dynamics can be made. It is found that, in all of these cases, the instantaneous vorticity-magnetic field contains a spectrum of numerous magnetized vorticity eddies of different scale and strength. The spatio-temporal evolution of the combined field is a strong function of initial data. In particular, if a strong Lorentz force is present, the resulting equations contain a cubic non-linearity in the current density. In comparison with the Navier-Stokes dynamics, the flow development is much more complex due to the mutual influence of viscous and Ohmic dissipations. 

\end{abstract}

\section{Introduction}
We intend to study the global regularity of magnetohydrodynamics (or MHD), which describes the evolution of electrically conducting fluids in the presence or absence of body forces, electric currents, and electromagnetic fields. We consider only the viscous and incompressible fluids which are permeated by magnetic fields. Thus, the governing equations are the Navier-Stokes equations of motion and pre-Maxwell equations, coupled via the Lorentz force and Ohm's law. For the general theory of magnetohydrodynamics and its applications, consult the books by Shercliff 1965; Robert 1967; Moffatt 1983; Davidson 2001; Lorrain {\it et al} 2006.

The {\itshape local} well-posedness of the Cauchy problem for the MHD equations has been known for a long time. One can trace the original ideas back to the work of Leray and Hopf on the Navier-Stokes equations (Leray 1934; Hopf 1951). Duvaut and Lions (1972) constructed a global weak solution and a local strong solution to the Cauchy problem. Roughly speaking, a weak solution or a suitable weak solution refers to a solution of the problem which has been shown to exist without known uniqueness and regularity properties while a strong solution is well-posed in certain designated Lebesgue or Sobolev spaces. For Sobolev spaces $H^s$, it was established in Sermange \& Temam (1983) for any given initial data $(u_0, B_0) {\in} H^s(\real^3), s {\geq} 3$. However, the question of the {\itshape global} regularity is still an open problem for $s{<}3$. On the basis of the partial regularity for the Navier-Stokes dynamics (Scheffer 1976; Caffarelli {\it et al} 1982), a generalized result on the partial regularity for MHD is considered in He \& Xin (2005). Particular attention should be paid to the fact that the solutions are not regular enough to ensure the law of conservation of energy.
Analogous regularity criteria to the widely-quoted Serrin-Prodi-Ladyzhenskaya criteria in mathematical theory of the Navier-Stokes dynamics are available. The regularity criteria have been improved in Chen {\itshape et al} (2008), using the Fourier localization method and Bony's para-product decomposition. 

The existence of a total energy cascade and the mechanisms of the energy transfers between the velocity and magnetic fields are investigated (Bradshaw \& Gruji\'{c} 2012). Essentially, the results are derived by applying the concepts of the Richardson energy cascades and of vorticity stretching. The study shows that it is the properties of hydrodynamic turbulence which first need to be addressed satisfactorily.
The concept of an energy cascade in MHD turbulence is popular in the physics and fluid dynamics community, but there is considerable disagreement regarding the origin of the cascade. The debate is hardly new, and it is largely the same problem as in the Navier-Stokes dynamics. Due to the interaction between the magnetic and vorticity fields, energy can also be transferred between the two fields and hence there are a number of transfer mechanisms. This fact simply makes the problem even harder to resolve.
 
The main difficulty in the treatment of the MHD equations is that there has been a lack of {\itshape a priori} estimates for the fluid and magnetic fields. These estimates must be sufficiently strong to guarantee global regularity. 

In the present paper, we shall follow the traditional approach to the MHD equations, in which the motion is formulated as a system of evolution equations for the fluid velocity (or vorticity) and the magnetic field, along with an auxiliary equation for the electric field superimposed on the fluid. Because of the recent progress in the theory of incompressible turbulence in $\real^3$ (Lam 2013), we intend to make an effort to analyze the MHD turbulence in $\real^3$ which is clearly an idealized problem, as the unknown currents inducing the electric and magnetic fields must be practically carried by conductors of finite extent. Nevertheless, the MHD problem in $\real^3$ would avoid some theoretical difficulties in analysis, while it is a generic model in establishing the connection between the fluid turbulence and its interaction with the electric and magnetic fields. Our experience in obtaining the global well-posedness for the Navier-Stokes dynamics suggests that we adopt a ``vorticity-current formulation'' which gives rise to a system of evolution equations for the unknown fluid vorticity and the magnetic field. When the system is converted into a system of integral equations, the result offers us a convenient platform where we can establish certain important {\itshape a priori} bounds. The availability of these bounds is crucial for regularity and uniqueness. The incompressibility hypothesis implies that the density remains constant throughout the motion, so that any influence due to density stratification can only be assessed in a compressible MHD theory. We shall postpone our efforts on the effects of Coriolis and buoyancy forces to subsequent studies. In addition, we shall not go into the details of whether a turbulent regime is physically reasonable for applications, such as the Solar wind and the interstellar plasmas. Rather, we shall concentrate on the idealized problem of incompressible flows, where the study is carried out to address the solutions of the MHD equations for finite, non-zero magnetic Prandtl number (the ratio of the fluid viscosity over the magnetic diffusivity). Consequently, we shall work with real fluids with positive magnetic diffusivity, so that both viscous and Ohmic dissipations are operative over the entire course of flow evolution.  

\section{The Cauchy problem}

\subsection{Governing equations}

We deal with flow evolution of a viscous, incompressible, electrically conducting fluid, in the whole space $\real^3$, in the absence of a prescribed force. We are interested in the interaction of the forces of fluid dynamics, electric currents, and electromagnetic fields. In principle, the fluid flow is governed by the Navier-Stokes equations, coupled to Maxwell's equations. We seek to solve a system of evolution equations for the fluid velocity $u$ and the magnetic field $B$ in the fluid; both are solenoidal vector fields, depending on time $t \in [0, T]$ and position $x \in \real^3$. The time $T$ is given and is assumed to be finite. The coupled system has the form
\begin{equation} \label{eq:mhd}
 \begin{split}
	{\partial u}/{\partial t} - \nu \Delta u &= -  (u . \nabla) u - \nabla p / \rho + (J \times B) /\rho, \\
	{\partial B}/{\partial t} - \lambda \Delta B &=   (B. \nabla) u  - (u . \nabla) B, \\ 
	\nabla.u & =0,\\
	\nabla.B & =0.  
	\end{split}
\end{equation}
The Laplacian is denoted by $\Delta$. The symbol $p$ is the pressure, the letter $\rho$ the density of the fluid. The kinematic viscosity and the magnetic diffusivity are denoted by $\nu$ and $\lambda\;(=1/(\mu_0 \sigma))$ respectively. (The electrical conductivity $\sigma$ is a given positive constant. The symbol $\mu_0$ stands for the permeability of free space and is a universal constant.)
The vorticity and the total current density are obtained by taking the curl of the velocity and the magnetic field: 
\begin{equation} \label{eq:curl-vm-field}
\omega = \nabla \times u,  \;\;\; \mbox{and} \;\;\;  {\mu_0} J = \nabla \times B.
\end{equation}
The vorticity is a familiar fluid dynamics characteristic. The second relation is known as Amp\'{e}re's law; it is a simplified version of the Amp\'{e}re-Maxwell equation, where the effect of Maxwell's correction for the displacement current has been neglected. The current density $J$ in the fluid obeys Ohm's law
\begin{equation} \label{eq:ohm}
	J = \sigma \big( E + u \times  B \big),
\end{equation}
where $E$ denotes the (unknown) electric field. In a conducting fluid, we use an extended Ohm's law to describe the total electromagnetic force per unit charge. The Lorentz force (per unit volume),
\begin{equation} \label{eq:lorentz}
	J \times B,
\end{equation}
arises from the charge interaction in a magnetic field. It acts on all conductors carrying a current in the field. Faraday's law states that the electric field is generated by the rate of change of the magnetic field:
\begin{equation} \label{eq:electric-field}
	\nabla \times E = - \partial B / \partial t.
\end{equation}
The first equation in (\ref{eq:mhd}) is just Newton's second law of motion. Compared to the Navier-Stokes momentum equation, the Lorentz force acts on the current conductor and restrains the relative movement of the magnetic field and the fluid. Its sign is opposite to the positive pressure gradient and appears to be a retarding force. The second equation is known as the induction equation. It is derived by taking the curl of Ohm's law (\ref{eq:ohm}), and taking into account the relation in (\ref{eq:electric-field}). The third equation in (\ref{eq:mhd}) is the continuity equation; it is a consequence of the incompressibility hypothesis. The last equation, the magnetic field being solenoidal, is implied in Faraday's law in view of the vector identity $\nabla.\nabla{\times}E{=}0$. Furthermore, these solenoidal conditions assert that
\begin{equation} \label{eq:continuity}
\nabla. \omega = 0, \;\;\; \mbox{and} \;\;\; \nabla. J = 0.
\end{equation}

The initial velocity for the Cauchy problem is given by
\begin{equation} \label{eq:vel-ic}
 u (x,0) = u_0 (x) \;\; (\in C^{\infty}(\real^3)),\;\;\; x \in \real^3,
\end{equation} 
and the initial magnetic field is specified as
\begin{equation} \label{eq:mag-ic}
 B (x,0) = B_0 (x) \;\; (\in C^{\infty}(\real^3)),\;\;\; x \in \real^3.
\end{equation} 
We are interested in the global regular solutions of the Cauchy problem for the coupled system in $\real^3$. To facilitate our analysis, we consider the initial data as localized:
\begin{equation} \label{eq:ic-localization}
 \begin{split}
	\big\| \: \big(1 + | x | \big)^{k_0} & \: \partial_x^{\alpha_0} u_0(x) \: \big\|_{L^{\infty}(\real^3)} < \infty,\\ 
	\big\| \: \big(1 + | x | \big)^{k_0} & \: \partial_x^{\alpha_0} B_0(x) \: \big\|_{L^{\infty}(\real^3)} < \infty. 
 \end{split}
\end{equation}
for any values of $k_0$ and $\alpha_0$. 

Since we are working in $\real^3$, the total energy may appear to be infinite for an observer moving with a constant finite speed along any straight path which can be arbitrarily far away from the origin. It is defensible that the energy appears to be unbounded. Such an anomaly is due to our choice of frame of reference. However, we can reformulate our magneto-fluid dynamics problem by choosing a Galilean transform, so that the observer becomes stationary relative the motion. Thus, the energy will remain finite unless the velocity is out of bound during the flow development. In subsequent analysis, we take it for granted that a Galilean transform is effected. 

Most mathematics symbols and notations used in the present work are standard. The parabolic cylinder in space-time $(x,t)$ is written as $\real^3 {\times} [0,t]$ for given fixed $0{<}t{\leq}T$. We use a multi-index notation for the spatial partial derivatives of order $\alpha$
\begin{equation*}
	\partial_x^{\alpha} f(x) = {\partial ^{\alpha} f(x) } / ({\partial x_1^{\alpha_1} {\cdots} \partial x_m^{\alpha_m}  } )
\end{equation*}
for all multi-indexes $\alpha{=}(\alpha_1, \cdots, \alpha_m)$ with $|\alpha|{=}\alpha_1 + \cdots + \alpha_m$.
All $\alpha$'s are non-negative integers. For integrals over $\real^3$, we write $\int(\cdot)\rd x$ for $\int_{\real^3}(\cdot)\rd x$. Let the Newtonian potential ${\cal N} (x){=} {(4 \pi)^{-1}}{|{x}|^{-1}}$. We use the convolution notation
\begin{equation*}
	{\cal K}{*}g(x) = \frac{1}{4 \pi}\int\nabla_x\Big( \frac{1}{ |x{-}y|}\Big) \times g(y) \rd y.
\end{equation*}

\subsection{Derivation of a priori bounds}

The first equation in (\ref{eq:mhd}) can be rewritten as
\begin{equation} \label{eq:mhd-mod}
\begin{split}
  {\partial u}/{\partial t} - \nu \Delta u & =  - (u.\nabla)u + (B. \nabla)B/(\rho \mu_0) -  \nabla \big(  p {+} {B^2}/{(2 \mu_0)} \big)/\rho\\ 
	& =  u \times \omega + (J \times B) /\rho + \nabla \chi, 
\end{split}
\end{equation}
where $\chi = -(\nabla p/\rho + u^2/2)$ is the Bernoulli-Euler pressure. The last term on the right in the first identity shows that the quantity $B^2$ acts on the fluid as if it is a pressure. As it is irrotational, hence it does not contribute to the vorticity. 

To allow for possible singularities in the velocity and in the vorticity during the evolution of flow, we proceed our analysis by using a standard mathematical device. We introduce a sequence of regularizations of the solutions of the vorticity and magnetic equations (see, for example, Caffarelli {\it et al} 1982). For fixed $N{>}0$, we denote the regularizations by $\delta{=}T/N$. The symbol $\delta$ is known as the mollification parameter. We introduce the notation $\phi_{\delta}(u)$ to denote the mollification of the velocity $u$ in space and in time:
\begin{equation*}
	 \phi_{\delta}(u) (x,t) = \delta^{-4} \int_{\real}  \int_{\real^3} {\widetilde \phi} \Big( \frac{y}{\delta}, \frac{s}{\delta} \Big) \: u (x{-}y,t{-}s) \rd y \rd s.
\end{equation*}
The kernel function, $ {\widetilde \phi}(x,t) {\in} C^{\infty}_c(1{<}t{<}2,|x|{<}1)$, is a mollifier which is a smooth function in space and in time with compact supports. In addition, it has the properties: ${\widetilde \phi}{\geq}0$, and 
\begin{equation*}
	\int_{\real} \int_{\real^3}  {\widetilde \phi}(x,t) \rd x \rd t =  1.
\end{equation*}

The structures of the second equation in (\ref{eq:mhd}) and momentum equation (\ref{eq:mhd-mod}) suggest that we only need to mollify the velocity such that
\begin{equation} \label{eq:mhd-mollified}
 \begin{split}
 {\partial_t u} - \nu \Delta u & =  \phi_{\delta}(u) \times (\nabla \times u) + (J \times B) /\rho + \nabla \chi, \\
 {\partial_t B} - \lambda \Delta B & =   \nabla \times ( \phi_{\delta}(u) \times  B). 
	\end{split}
\end{equation}
We have omitted the mollification on the velocity continuity, as the mollification preserves the solenoidal property of $u$, in view of the well-known properties of mollifiers (see, for example, Adams \& Fournier 2003). 

We take the curl on the momentum equation, and we find that the system reduces to
\begin{equation} \label{eq:curls-mollified}
 \begin{split}
	\partial_t \omega_i - \nu \Delta \omega_i = \Big( (\omega . \nabla) \phi_{\delta}(u_i)  {-} (\phi_{\delta}(u) . \nabla )\omega_i \Big)  + \Big( \nabla {\times} (J {\times} B)/\rho \Big) & = {\mathscr R}_i, \\
	\partial_t B_i - \lambda \Delta B_i = (B. \nabla) \phi_{\delta}(u_i) {-} (\phi_{\delta}(u) . \nabla) B_i & = {\mathscr Q}_i.
	\end{split}
\end{equation}
The two terms in the middle equality of the first equation are the couple contributions from the velocity and magnetic fields. The signs of the couples are dynamically dependent on the local gradients of the velocity and the Lorentz force. From the principle of conservation of angular momentum, the presence of the magnetic field is equivalent to a prescribed non-conservative force. Hence it exerts an additional torque on the vorticity evolution.

First we observe that if we integrate the right-hand sides in (\ref{eq:curls-mollified}) over space $\real^3$, and then carry out integration by parts, we obtain the following summability conditions
\begin{equation} \label{eq:rhs-integrability}
	\sum_{i=1}^{3} \int {\mathscr R}_i(x,t) \rd x =0\;\;\; \mbox{and} \;\;\; \sum_{i=1}^{3} \int {\mathscr Q}_i(x,t) \rd x =0.
\end{equation}
It is evident that these relations are independent of the mollification parameter $\delta$. This observation suggests that it is advantageous to work in terms of the total vorticity and the total magnetic field
\begin{equation*}
	\omega_1 + \omega_2 + \omega_3 = \Omega, \;\;\; \mbox{and} \;\;\; B_1 + B_2 + B_3 = \Pi.
\end{equation*}
Hence the totals are governed by equations
\begin{equation} \label{eq:curls-total}
	\partial_t \Omega - \nu \Delta \Omega = \sum_{i=1}^3 {\mathscr R}_i(x,t,u,B)\;\;\; \mbox{and} \;\;\;
	\partial_t \Pi - \lambda \Delta \Pi =  \sum_{i=1}^3 {\mathscr Q}_i(x,t,B,u)
\end{equation}
respectively. By Duhamel's principle, these total quantities satisfy scalar integral equations
\begin{equation} \label{eq:vort-int}
	\Omega(x,t) = \int Z(x-y,t) \Omega_0(y) \rd y +  \int_0^{t} \!\! \int Z(x,t,y,s) \sum_{i=1}^{3}  {\mathscr R}_i(y,s)  \rd y \rd s
\end{equation}
and
\begin{equation} \label{eq:mag-int}
	\Pi(x,t) = \int Y(x-y,t) \Pi_0(y) \rd y +  \int_0^{t} \!\! \int Y(x,t,y,s) \sum_{i=1}^{3}  {\mathscr Q}_i(y,s)  \rd y \rd s.
\end{equation}	
The notations, $\Omega_0$ and $\Pi_0$, stand for the total initial data respectively. The integral kernels $Z$ and $Y$ are the fundamental solutions of diffusion equation and are defined by
\begin{equation} \label{eq:heat-kernel}
	Z(x,t,y,s) = \big( 4 \pi \nu (t{-}s) \big)^{-3/2} \exp \Big( - \frac{|x{-}y|^2} {4 \nu (t{-}s) } \Big), \;\;\;{t>s},
\end{equation}
and
\begin{equation} \label{eq:diff-kernel}
Y(x,t,y,s) = \big( 4 \pi \lambda (t{-}s) \big)^{-3/2} \exp \Big( - \frac{|x{-}y|^2} {4 \lambda (t{-}s) } \Big), \;\;\;{t>s}.
\end{equation}

A further integration on (\ref{eq:mag-int}) and trivial applications of the Fubini's theorem enable us to infer the invariance of the total magnetic field. The same applies to the total vorticity. Thus we assert that
\begin{equation} \label{eq:vort-mag-int}
\frac{\rd }{\rd t} \int \Omega(x,t) \rd x = 0, \;\;\; \mbox{and} \;\;\;
\frac{\rd }{\rd t} \int \Pi(x,t) \rd x = 0
\end{equation}
by virtue of the well-known properties of the fundamental solutions for diffusion.
 
It follows that the first invariance implies 
\begin{equation} 
	\int \Omega(x,t) \rd x = \int \bar{\omega}_0(x) \rd x \;{\leq} \;  \| \Omega_0\|_{L^1 (\real^3)}.
\end{equation}
Consider $\bar{\omega}$ as a measurable, real-valued function. We write $\Omega = \Omega^{+}  - \Omega^{-}$, where both $\Omega^{+}{=}\mbox{max}(\Omega,0)$ and $\Omega^{-} {=}{-}\mbox{min}(\Omega,0)$ are measurable, non-negative, and {\itshape{finite}}. Hence the set $\{\Omega {=} {+} {\infty}\} \cup\{\Omega {=} {-} {\infty}\}$ has measure zero. By the Archimedean property of the real numbers (Royden \& Fitzpatrick 2010), we deduce that every component of vorticity satisfies the bound
\begin{equation*} 
	\big\|\Omega_i(x)\big\|_{L^1 (\real^3)} \;{\leq}\; {\mathbb N} \; \big\|\Omega_0(x)\big\|_{L^1 (\real^3)},
\end{equation*}
which holds, except possibly on a set of measure zero. The symbol ${\mathbb N}{=}{\mathbb N}(T)$ denotes a natural number and it can be suitably chosen for any $t {\in} [0,T]$. Using the elementary identity
\begin{equation*}
	\omega_1^2 + \omega_2^2 + \omega_3^2 \: {\leq} \: \big( |\omega_1^2| + |\omega_2^2| + |\omega_3^2| \big)^2,
\end{equation*}
and the fact that $|(\omega_0)_i|{\leq}|\omega_0|$, the integrability bound can be upgraded to
\begin{equation} \label{eq:vort-l1-norm}
	\big\|\omega(x)\big\|_{L^1 (\real^3)} \;{\leq}\; {\mathbb N} \; \big\|\omega_0(x)\big\|_{L^1 (\real^3)}
\end{equation}
for some natural number ${\mathbb N}$. By analogy, we have
\begin{equation} \label{eq:mag-l1-norm}
	\|B(x)\|_{L^1 (\real^3)} \;{\leq}\; {\mathbb N_1} \; \|B_0\|_{L^1 (\real^3)}.
\end{equation}

In fact, we can generalize this procedure as follows. For any integers $\alpha{\geq}0$ and $\beta{\geq}0$, we derive the equations
\begin{equation} \label{eq:mollified-d-vort}
	\partial_t \: (\partial_t^{\beta} \partial_x^{\alpha} \omega_i) - \nu \Delta (\partial_t^{\beta}\partial_x^{\alpha} \omega_i) = \partial_t^{\beta}\partial_x^{\alpha} {\mathscr R}_i(x,t),
\end{equation}
and
\begin{equation} \label{eq:mollified-d-mag}
	\partial_t \: (\partial_t^{\beta}\partial_x^{\alpha} B_i) - \lambda \Delta (\partial_t^{\beta}\partial_x^{\alpha} B_i) = \partial_t^{\beta} \partial_x^{\alpha} {\mathscr Q}_i(x,t).
\end{equation}
In view of the solenoidal conditions in (\ref{eq:mhd}) and (\ref{eq:continuity}), we verify that
\begin{equation} 
	\sum_{i=1}^3 \int \partial_t^{\beta} \partial_x^{\alpha} {\mathscr R}_i \rd x = 0
\end{equation}
because the operators $\partial_t^{\beta}\partial_x^{\alpha}$ and $\nabla$ commute. Similarly, we arrive at
\begin{equation} 
	\sum_{i=1}^3 \int \partial_t^{\beta} \partial_x^{\alpha} {\mathscr Q}_i \rd x = 0.
\end{equation}
Thus we establish the generalization of the invariants in (\ref{eq:vort-mag-int}):
\begin{equation} \label{eq:vm-int}
\frac{\rd }{\rd t} \int \partial_t^{\beta} \partial_x^{\alpha} {\Omega}(x,t) \rd x = 0, \;\;\; \mbox{and} \;\;\; \frac{\rd }{\rd t} \int \partial_t^{\beta} \partial_x^{\alpha} \Pi(x,t) \rd x = 0.
\end{equation}
From these integral constraints and the Sobolev embedding theorem (see, for example, Adams \& Fournier 2003), we conclude that
\begin{equation} \label{eq:smooth-ap-bound}
	\omega \in C_B^{\infty}, \;\;\; \mbox{and} \;\;\; B \in C_B^{\infty}.
\end{equation}
Evidently, these remarkable {\itshape a priori} bounds are independent of the mollifications. 
Thus we can work in the space of smooth bounded functions for the vorticity and magnetic fields in $\real^3{\times}[0,T]$ in our subsequent analysis. In fact, the first bound also implies that, {\itshape a priori}, $u \in C_B^{\infty}$.

We remark that the solenoidal conditions, $\nabla.u{=}0$ and $\nabla.B{=}0$, define a rather weak constraint on the velocity and the magnetic field, as they suggest that every term in $\partial u_i/\partial x_i$, and in $\partial B_i/\partial x_i$ is finite. 

Consequently, the Cauchy problem for magnetohydrodynamics is defined in the following coupled equations governing the vorticity and magnetic fields  
\begin{equation} \label{eq:curls}
 \begin{split}
	\partial_t \omega_i - \nu \Delta \omega_i & = (\omega . \nabla) u_i  - (u . \nabla )\omega_i  + \Big((B . \nabla) J_i  - (J . \nabla )B_i \Big)/\rho, \\
	\partial_t B_i - \lambda \Delta B_i & =   (B. \nabla) u_i - (u . \nabla) B_i.
	\end{split}
\end{equation}
Either field is solenoidal. The initial vorticity for the problem is given by
\begin{equation} \label{eq:vort-ic}
 \omega (x,0) = \nabla {\times} u_0 (x) = \omega_0 (x),
\end{equation} 
and the initial current density is specified as
\begin{equation} \label{eq:current-ic}
 J (x,0) = \nabla {\times} B_0 (x)/\mu_0 = J_0 (x).
\end{equation} 
In addition, the initial data are considered as solenoidal and localized.

Thus the well-posedness of the Cauchy problem for the magnetohydrodynamics equations can be completely established according to the well-developed theory of parabolic systems with bounded smooth coefficients (see, for example, Ladyzhenskaya {\itshape et al} 1968; Friedman 1964; Eidel'man 1969). 

The following Poisson's equation defines the kinematic relationship between the solenoidal field $B$ and its curl:  $\Delta \Psi(x) {=} {-}  {\mu}_0 J(x)$,
where $\Psi$ denotes a vector function (see, for example, Lorrain \& Corson 1970).
The magnetic field is the distributional derivative of the vector and it is given by the Biot-Savart law
\begin{equation} \label{eq:bs}
B(x)=\nabla \times \Psi(x) = \frac{{\mu}_0}{4 \pi} \int \frac{(x-y)}{|x-y|^3} \times J(y) \rd y = {\mu_0}{\cal K}{*}J(x)
\end{equation}
at any instant of time. 
The Biot-Savart law also holds for the velocity field, we give the explicit form as
\begin{equation} \label{eq:bs-vel}
u(x)= \frac{1}{4 \pi} \int \frac{(x-y)}{|x-y|^3} \times \omega(y) \rd y = {\cal K}{*}\omega(x).	
\end{equation}

\subsection{Generalization to include the Hall-current effect}

Lighthill questioned the validity of the induction equation in (\ref{eq:mhd}) for certain applications in astrophysics (Lighthill 1960) where the main interest lies in the evolution of a plasma of electrons and positive ions. For real conducting fluids, the induction equation takes a modified form
\begin{equation} \label{eq:hmhd}
	{\partial B}/{\partial t} - \lambda \Delta B =   \nabla \times(u \times B) + \theta_e \nabla \times(J \times B),
\end{equation}
where the last term incorporates the so-called Hall current effect, and $\theta_e$ denotes the inverse of the electron charge density. In high-conductivity plasmas, the extra term or the Hall term is considered to be responsible for the phenomena of magnetic reconnection that involve redistribution of magnetic energy and kinetic energy. Thus the rate of change of $J{\times}B$ plays a significant role in the evolution of the magnetic vorticity ($J$). On the basis of the mollifications, the induction equation now has a mollified form:
\begin{equation*} 
	\partial_t B_k - \lambda \Delta B_k = (B. \nabla) \phi_{\delta}(u_k) {-} (\phi_{\delta}(u) . \nabla) B_k + \theta_e \Big( (B. \nabla)J_k - (J.\nabla)B_k \Big),
\end{equation*}
where $\nabla.B{=}0$, and $k{=}1,2,3$.
Evidently, the sum of the right-hand under the spatial integration remains an invariant. Hence we can generalize our {\itshape a priori} bound (\ref{eq:smooth-ap-bound}) to include the Hall current effect.

However, a plasma consisting of electrons (whose mass and number density are denoted by $m_e$ and $n_e$ respectively) and ions ($m_i$ and $n_i$), the momenta of these two ``fluids'' take rather distinct values in space-time because their density ratio $\rho_i/\rho_e = (n_i m_i) /(n_e m_e)  \approx 1800$. Consequently, the velocity $u$ in the hydrodynamic equations (\ref{eq:mhd}) needs to be specified to allow for the momentum differentials. It is clear that the velocity of the two combined streams is actually closer to $v_i$, the mean velocity of the ions, than to $v_e$, that of the electrons. This fact calls for better approximations in the equations of motion. The (vector) vorticity equations and the induction equation are written separately for the electrons, the ions, and the magnetic field as follows:
\begin{equation} \label{eq:sep-curls}
 \begin{split}
	\partial_t \omega_e - \nu_{e} \Delta \omega_e & = - \nabla {\times} (v_e {\times} \omega_e) - (\vartheta_e/\rho_e)\:\nabla {\times} (J {\times} B) - \vartheta_e \nabla {\times} (v_e {\times} B) - {\mathscr W} /\rho_e, \\
	\partial_t \omega_i - \nu_{i} \Delta \omega_i & = - \nabla {\times} (v_i {\times} \omega_i) + (\vartheta_i/\rho_i)\: \nabla {\times} (J {\times} B) + \vartheta_i \nabla {\times} (v_i {\times} B) + {\mathscr W}/\rho_i,\\
	\partial_t B - \lambda \Delta B & = - \nabla {\times} (v_e {\times} B) + \theta_e \nabla {\times} (J {\times} B),
	\end{split}
\end{equation}
where the coefficients of viscosity for the electrons ($\nu_e$) and the ions ($\nu_i$) take different values. The parameters, $\vartheta_i=\vartheta_i(n_i,n_e,m_i)$ and $\vartheta_e=\vartheta_e(n_i,n_e,m_e)$, depend also on the electron charge.
Because of collisions with ions, there is a loss of electron momentum per unit volume. We denote the curl of the rate of the normalized loss by ${\mathscr W}$, and 
\begin{equation*}
	{\mathscr W} = - \nabla {\times} J. 
\end{equation*}
As just discussed, the stream velocity is far greater than $v_e$; the term involving $v_e$ in the induction equation is bound to be small compared to the Hall term. The third equation in (\ref{eq:sep-curls}) defines an invariance of the magnetic field, i.e., the second integral relation in (\ref{eq:vm-int}). In particular, we have
\begin{equation*}
	\frac{\rd }{\rd t} \int \Delta \Pi(x,t) \rd x = 0.
\end{equation*}
From the first two differential equations, we assert that the following {\itshape a priori} bounds hold
\begin{equation} \label{eq:sep-vort-int}
\begin{split}
\rho_e \frac{\rd }{\rd t} \int {\Omega}_e(x,t) \rd x & =- \rho_i \frac{\rd }{\rd t} \int {\Omega}_i(x,t) \rd x \\
&= - \frac{1}{\mu_0} \int \Delta \Pi(x,0) \rd x,
\end{split}
\end{equation}
where $\Omega_e$ and $\Omega_i$ stand for the total vorticity for electrons and ions respectively. The quantity of the last integral is completely specified by the initial data. It is clear how to compute the integrals of $\partial_t^{\beta} \partial_x^{\alpha} {\Omega}_e$ for $\alpha,\beta{\geq}0$.
 
\section{Weak Lorentz force}

If the magnetic field is so weak that its interaction with the vorticity field can be neglected, then the fluid dynamics equations and the induction equation in MHD system (\ref{eq:mhd}) become uncoupled. Equations (\ref{eq:curls}) reduce to the approximation
\begin{equation} \label{eq:curls-approx}
 \begin{split}
	\partial_t \omega_i - \nu \Delta \omega_i & = (\omega . \nabla) u_i  - (u . \nabla )\omega_i, \\
	\partial_t B_i - \lambda \Delta B_i & =   (B. \nabla) u_i - (u . \nabla) B_i.
	\end{split}
\end{equation}
These equations are to be solved subject to the initial conditions (\ref{eq:vort-ic}) and (\ref{eq:current-ic}). It is evident that the role of the fluid dynamics is crucial in the global regularity of the incompressible MHD equations in $\real^3$, since the velocity becomes a coefficient function in the induction equation.

The first equation in (\ref{eq:curls-approx}) can be converted into an integral equation:
\begin{equation} \label{eq:vort-ie}
	\omega_i(x,t) = f_i(x,t) + \int_0^{t} \int Z(x,t,y,s) \Big( (\omega . \nabla) u_{i}  - (u . \nabla )\omega_{i}\Big) (y,s)  \rd y \rd s, 
\end{equation}
where $f_i$ denotes the components of the vector function $f$, which is given by
\begin{equation} \label{eq:caloric-vort-ic}
	f(x,t) = \int \mathbf{Z}(x{-}y,t) \: \omega_0(y) \rd y.
\end{equation}
The matrix $\mathbf{Z}$ is diagonal. Its elements are identical and equal to $Z$. Similarly,
\begin{equation} \label{eq:mag-ie}
	B_i(x,t) = b_i(x,t) +  \int_0^{t} \int Y(x,t,y,s) \Big( (B . \nabla) u_{i}  - (u . \nabla )B_i\Big) (y,s)  \rd y \rd s, 
\end{equation}
where $b$ is the mollified initial magnetic field 
\begin{equation} \label{eq:caloric-mag-ic}
	b(x,t) = \int \mathbf{Y}(x{-}y,t)  B_0(y) \rd y.
\end{equation}
The matrix $\mathbf{Y}$ is diagonal. Its elements are identical and equal to $Y$.
Equations (\ref{eq:vort-ie}) and (\ref{eq:mag-ie}) define the vorticity and the current density as distribution solutions. 

The global well-posedness for the Navier-Stokes equations or for the vorticity equation has been established in Lam (2013). In particular, the vorticity has a series solution:
\begin{equation} \label{eq:vort-series-sol}
 \begin{split}
	\omega(x,t) = \gamma(x,t) \: & + \sum_{m{\geq}2}\: \Big( \sum_{k=1}^{S_m} \: V_k[\gamma]^m\Big) = \gamma(x,t) \: + \sum_{m{\geq}2}\: {S_m} \: V[\gamma]^m  \\
	\quad & \\
	= \gamma(x,t) \: & + 2 \: V[\gamma]^2 + 10 \: V[\gamma]^3 + 62 \: V[\gamma]^4 + 430 \: V[\gamma]^5 + 3194 \: V[\gamma]^6  \\ 
	 \quad & \\
	\quad & + 24850 \: V[\gamma]^7 + 199910 \: V[\gamma]^8 +  \cdots \;\;\; \big(= {\mathscr L}\big( \gamma \big) \big),
 \end{split}
\end{equation}
where $\gamma$ is the solution of the Volterra-Fredholm integral equation
\begin{equation*} 
	\gamma(x,t) = f(x,t) + \int_0^{t} \!\! \int K(x,t,y,s) \gamma(y,s) \rd y \rd s.
\end{equation*}
The kernel $K$ is the mollified kernel $G$ by the initial vorticity. The elements of $G$ are defined by
\begin{equation*}
	K_{ij}(x,t,y,s)= \int G_{ij}(x,t,y,s,z) f(z) \rd z.
\end{equation*}
We use the notation $G_{ij}$ to denote the row matrix 
\begin{equation*}
	G_{ij}=\big( \; \alpha_{ij} \;\;\; \beta_{ij} \;\;\; \gamma_{ij} \; \big).
\end{equation*}
The elements for $G$ are the product of the derivative of the heat kernel $Z$ and the derivative of the Newtonian potential originated from the Biot-Savart formula. It is convenient to introduce some abbreviations:
\begin{equation} \label{eq:abbrev}
	Z_i = Z_i(x,\tau,y,s) = \frac{\partial Z(x,\tau,y,s)}{\partial y_i}, \;\;\; N_i = N_i(y,z) = \frac{\partial {\cal N}(y{-}z)}{\partial y_i},\;\;i{=}1,2,3,
\end{equation} 
where the spatial differentiations for the Newtonian potential ${\cal N}$ are taken at every instant of time $s\leq T$.
In terms of these notations, the elements of $G_{ij}$ are given by
\begin{equation} \label{eq:gij}
\left. \begin{aligned} 
	{\alpha}_{11} & = Z_2 N_3-Z_3N_2 & {\alpha}_{12} & = 0, & {\alpha}_{13} & = 0, \\
	{\beta}_{11}  & = Z_3N_1, & {\beta}_{12} & = Z_2N_3, & {\beta}_{13}    & =Z_3N_3, \\
	{\gamma}_{11} & = -Z_2N_1,  & {\gamma}_{12}& = -Z_2N_2,  & {\gamma}_{13}   &= -Z_3N_2, \\
	& & & & & \\
	{\alpha}_{21} & = -Z_1N_3  & {\alpha}_{22}& = -Z_3N_2,  & {\alpha}_{23} & = -Z_3N_3, \\
	{\beta}_{21}  & = 0,       & {\beta}_{22} & = Z_3N_1-Z_1N_3, & {\beta}_{23}  & = 0, \\
	{\gamma}_{21} & = Z_1N_1,  & {\gamma}_{22}& = Z_1N_2,  & {\gamma}_{23} &= Z_3N_1, \\ 
	& & & & & \\
	{\alpha}_{31} & = Z_1N_2  & {\alpha}_{32}& = Z_2N_2,  & {\alpha}_{33} & = Z_2N_3, \\
	{\beta}_{31}  & = -Z_1N_1,  & {\beta}_{32} & = -Z_2N_1,   & {\beta}_{33}  & = -Z_1N_3, \\
	{\gamma}_{31} & = 0,       & {\gamma}_{32}& = 0,  & {\gamma}_{33} &= Z_1N_2-Z_2N_1. 
 \end{aligned}
 \right \} 
\end{equation}
In (\ref{eq:vort-series-sol}), the notations, $V[\gamma]^m$, are used to present an integro-power term of degree $m$ with respect to function $\gamma$:
\begin{equation*}
  \begin{split}
\idotsint	{\mathrm K}(x,t,x_i,t_i,{\cdots},x_2,t_2,x_1,t_1) &\gamma^{k_1}(x_1,t_1)\gamma^{k_2}(x_2,t_2) {\cdots} \gamma^{k_i}(x_i,t_i) \\
	\quad & \rd x_1 \rd t_1 \rd x_2 \rd t_2 {\cdots} \rd x_i \rd t_i,
	\end{split}
\end{equation*}
where $k_1{+}k_2{+}\cdots{+}k_i{=}m$. The integers, $S_m$, are the numbers of allowable combinations of the vorticity $\gamma$. In summary, every term in the series expansion (\ref{eq:vort-series-sol}) constitutes a space-time convoluted vorticity. Each constituent is called an eddy. 
Evidently, the initial vorticity lies in the core of every eddy. Furthermore, the entire vorticity field is populated by the vorticity integro-power terms of all the degrees. This intricate vorticity field is the general solution of the vorticity equation and it is turbulence.
 
The integral equation in (\ref{eq:mag-ie}) for the magnetic field can be rewritten in the following vector form
\begin{equation} \label{eq:mag-ie-approx}
	B(x,t) = b(x,t) + \int_0^{t} \!\! \int   U (x,t,y,s) B(y,s)  \rd y \rd s,
\end{equation}
where the kernel $U$ denotes a $3{\times}3$ matrix: 
\begin{equation} \label{eq:mag-ie-kernel}
	U(x,t,y,s)= \left( \begin{array}{ccc}
\nabla_{y}Y.u {+} Y \partial_{y_1}u_1 & Y \partial_{y_2}u_1  & Y \partial_{y_3}u_1 \\
Y \partial_{y_1}u_2 &  \nabla_{y}Y.u {+} Y \partial_{y_2}u_2 & Y \partial_{y_3}u_2 \\
Y \partial_{y_1}u_3 & Y \partial_{y_2}u_3 &  \nabla_{y}Y.u {+} Y \partial_{y_3}u_3 \\
\end{array} 
\right),
\end{equation}
and $u{=}u(y,s)$. Hence the interaction between the vorticity and magnetic fields in the absence of Lorentz force can be completely determined by evaluating the expression
\begin{equation} \label{eq:mag-ie-approx-sol}
	B(x,t) = b(x,t)  + \int_0^{t} \!\! \int  U^*(x,t,y,s) b(y,s) \rd y \rd s,
\end{equation}
where $U^*$ is the resolvent kernel of $U$. In view of (\ref{eq:bs}), $B$ is clearly solenoidal. In conclusion, turbulence is exclusively generated by the vorticity and is relayed into the magnetic field. This finding is contrary to a long-held view that random stretching of magnetic lines of force causes a systematic decrease in scale of the magnetic field and hence leads to an increase of magnetic energy density, as long as Ohmic dissipation can be neglected (see, for example, Batchelor 1950). The essence is that the field $B$ is not subject to the curl constraint $\omega {=} \nabla{\times}u$ and hence the non-linearity itself on the ``forcing'' side in the induction is {\itshape not} the appropriate type to proliferate small scale motions. 

\section{Iterative integral system}

\subsection{Existence}

We proceed to establish a sequence of approximations to the solutions of the coupled system (\ref{eq:curls}). The approximations are denoted by $\omega^{(k)}{=}\omega^{(k)}(x,t)$ and $B^{(k)}{=}B^{(k)}(x,t)$. Let $u^{(0)}{\equiv}0$. For $k=1,2,3,\cdots$, they are defined in the following iterative relations:
\begin{equation} \label{eq:iter-vort-mag}
 \begin{split}
 {\partial B^{(k)}}/{\partial t} - \lambda \Delta B^{(k)} & = (u^{(k-1)}.\nabla) B^{(k)} - (B^{(k)} . \nabla ) u^{(k-1)}, \\
	{\partial \omega^{(k)}}/{\partial t} - \nu \Delta \omega^{(k)} & = (\omega^{(k)}.\nabla) u^{(k)} - (u^{(k)} . \nabla ) \omega^{(k)} + {\mathscr F} \big( B^{(k)} \big), \\
	u^{(k)} (x)& = {\cal K} {*} \omega^{(k)}(x), \\ 
	B^{(k)}(x,0)& =B_0(x),\\
	\omega^{(k)}(x,0)& =\nabla{\times}u_0(x). 
 \end{split}
\end{equation}
The components of the function ${\mathscr F}{=}{\mathscr F}(x,t,B^{(k)})$ are calculated from the magnetic field
\begin{equation}
	{\mathscr F}_i \big( B^{(k)} \big) = \Big((B^{(k)} . \nabla) {(\nabla{\times}B^{(k)})_i}  - ({\nabla{\times}B^{(k)} } . \nabla )B^{(k)}_i \Big)/\rho.
\end{equation}
We notice that the solenoidal conditions in (\ref{eq:mhd}) and (\ref{eq:continuity}) hold for all values of $k$. 

For $k{=}1$, the first equation $B^{(1)}$ is just the solution of the pure initial value problem for diffusion heat equation $\partial B^{(1)} / {\partial t} {-} \nu \Delta B^{(1)}{=}0$. It can be readily solved by method of Fourier transform. Explicitly, 
\begin{equation*} 
	B^{(1)}(x,t) = \int   {\mathbf Y} (x,t,y) \; B_0(y) \rd y.
\end{equation*}
Hence the solenoidal $B^{(1)}=\mu_0{\cal K}*J^{(1)}$.
Thus the second equation can be written as
\begin{equation} \label{eq:vort-k1}
	\omega^{(1)}_i(x,t) =  f^{(1)}_i(x,t) + 
	\int_0^{t} \!\! \int \Big(  \big( \omega^{(1)}_i u^{(1)} - u^{(1)}_i\omega^{(1)}  \big).\nabla_y\Big) Z(x,t,y,s) \: \rd y \rd s,
\end{equation}
where
\begin{equation} \label{eq:f1}
   f^{(1)}(x,t) = \int {\mathbf Z} (x,t,y) \omega_0(y) \rd y + \int_0^t \!\! \int {\mathbf Z}(x,t,y,s){\mathscr F} \big( B^{(1)} \big)(y,s)  \rd y \rd s.
\end{equation}
Let $\gamma^{(1)}$ be the solution of the integral equation
\begin{equation*} 
	\gamma^{(1)}(x,t) = f^{(1)}(x,t) + \int_0^{t} \!\! \int K(x,t,y,s) \gamma^{(1)}(y,s) \rd y \rd s.
\end{equation*}
Equation (\ref{eq:vort-k1}) can be completely solved, and the result is given by
\begin{equation} \label{eq:omega1}
	\omega^{(1)}(x,t)={\mathscr L}\big( \gamma^{(1)} \big).
\end{equation}
Hence the velocity $u^{(1)}$ is solved accordingly.
Consider
\begin{equation}
	{\partial B^{(2)}}/{\partial t} - \lambda \Delta B^{(2)} = (u^{(1)}.\nabla) B^{(2)} - (B^{(2)} . \nabla ) u^{(1)}.
\end{equation}
Thus the solution for $B^{(2)}$ can be found by solving the following integral equation
\begin{equation} \label{eq:mag-ie-k2}
	B^{(2)}(x,t) = b(x,t) + \int_0^{t} \!\! \int  U^{(1)} (x,t,y,s) B^{(2)}(y,s)  \rd y \rd s,
\end{equation}
where $U^{(1)}$ is the kernel $U$ in (\ref{eq:mag-ie-kernel}) with every $u$ replaced by $u^{(1)}$. It follows that
\begin{equation} \label{eq:vort-k2}
	\omega^{(2)}_i(x,t) =  f^{(2)}_i(x,t) + 
	\int_0^{t} \!\! \int \Big(  \big( \omega^{(2)}_i u^{(2)} - u^{(2)}_i\omega^{(2)}  \big).\nabla_y\Big) Z(x,t,y,s) \: \rd y \rd s,
\end{equation}
where
\begin{equation}
f^{(2)}(x,t) = \int {\mathbf Z} (x,t,y) \omega_0(y) \rd y + \int_0^t \!\! \int {\mathbf Z}(x,t,y,s){\mathscr F} \big( B^{(2)} \big)(y,s)  \rd y \rd s.
\end{equation}
Let $\gamma^{(2)}$ be the solution of the integral equation
\begin{equation*} 
	\gamma^{(2)}(x,t) = f^{(2)}(x,t) + \int_0^{t} \!\! \int K(x,t,y,s) \gamma^{(2)}(y,s) \rd y \rd s.
\end{equation*}
The solution for equation (\ref{eq:vort-k2}) is found to be
\begin{equation}
	\omega^{(2)}(x,t)={\mathscr L} \big( \gamma^{(2)} \big).
\end{equation}
Suppose that the solution $u^{(k)}$ is known. Then the induction equation for $k{+}1$ becomes
\begin{equation} \label{eq:b-kp1}
	{\partial B^{(k+1)}}/{\partial t} - \lambda \Delta B^{(k+1)} = (u^{(k)}.\nabla) B^{(k+1)} - (B^{(k+1)} . \nabla ) u^{(k)}.
\end{equation}
Thus the solution for $B^{(k+1)}$ can be found by solving the following integral equation
\begin{equation} \label{eq:mag-ie-kp1}
	B^{(k+1)}(x,t) = b(x,t) + \int_0^{t} \!\! \int  U^{(k)} (x,t,y,s) B^{(k+1)}(y,s)  \rd y \rd s,
\end{equation}
where $U^{(k)}$ is the kernel $U$ in (\ref{eq:mag-ie-kernel}) with every $u$ replaced by $u^{(k)}$. This is an integral equation of the Volterra-Fredholm type. The solution is given by
\begin{equation} \label{eq:mag-ie-kp1-sol}
	B^{(k+1)}(x,t) = b(x,t)  + \int_0^{t} \!\! \int  W^{(k)}(x,t,y,s) b(y,s) \rd y \rd s,
\end{equation}
where $W^{(k)}$ is the resolvent kernel of $U^{(k)}$. Evidently, $B^{(k+1)}(x)=\mu_0 {\cal K}*J^{(k+1)}(x)$ for every instant of time. Thus the vorticity is determined from
\begin{equation} \label{eq:vort-ie-kp1}
	\omega^{(k+1)}_i(x,t) =  f^{(k+1)}_i(x,t) + 
	\int_0^{t} \!\! \int \Big(  \big( \omega^{(k+1)}_i u^{(k+1)} - u^{(k+1)}_i\omega^{(k+1)}  \big).\nabla_y\Big) Z \: \rd y \rd s,
\end{equation}
where
\begin{equation} \label{eq:f-kp1}
f^{(k+1)}(x,t) = \int {\mathbf Z} (x,t,y) \omega_0(y) \rd y + \int_0^t \!\! \int {\mathbf Z}(x,t,y,s){\mathscr F} \big( B^{(k+1)} \big)(y,s)  \rd y \rd s.
\end{equation}
Let $\gamma^{(k+1)}$ be the solution of the integral equation
\begin{equation*} 
	\gamma^{(k+1)}(x,t) = f^{(k+1)}(x,t) + \int_0^{t} \!\! \int K(x,t,y,s) \gamma^{(k+1)}(y,s) \rd y \rd s.
\end{equation*}
The solution for equation (\ref{eq:vort-k2}) is found to be
\begin{equation} \label{eq:vort-kp1}
	\omega^{(k+1)}(x,t)={\mathscr L} \big( \gamma^{(k+1)} \big).
\end{equation}
Furthermore, the velocity $u^{(k+1)}$ is known. As $k \rightarrow \infty$, we have
\begin{equation*}
	B^{(k)}(x,t) \rightarrow B(x,t), \;\;\; \mbox{and}\;\;\; \omega^{(k)}(x,t) \rightarrow \omega(x,t),\;\;\; \forall (x,t) \in \real^3{\times}[0,T]
\end{equation*}
in view of the Arzel{\`{a}}-Ascoli theorem. The existence proof is completed. 

\subsection{Uniqueness}

An adjoint equation for the magnetic field can be calculated from the bilinear Lagrange-Green identity (Lam 2013). We find the adjoint in its component-wise form:
\begin{equation} \label{eq:mag-adj}
	\frac {\partial B^{\dagger}_i} {\partial t} + \lambda \Delta B^{\dagger}_i = u_j \frac{\partial B^{\dagger}_i}{\partial x_j} + \frac{\partial u_j }{\partial x_i} B^{\dagger}_j.
\end{equation}
The adjoint starting condition is specified as
\begin{equation*}
	B^{\dagger}(x,T)=B_0(x).
\end{equation*}

Let $\Lambda$ denote the fundamental solution to the second equation in (\ref{eq:curls}). Then
the generalized solution of the Cauchy problem has an integral representation 
\begin{equation} \label{eq:pde-mag-sol}
	B(x,t) = \int {\Lambda}(x,t,y) \; B_0(y) \rd y.
\end{equation}
In addition, every element of $\Lambda$ possesses the Gaussian upper and lower bounds
\begin{equation} \label{eq:gaussian-bounds}
	\frac{C_1} {(\lambda(t{-}s))^{3/2}} \exp \Big( {-} C_2 \frac { |x{-}y|^2}{ \lambda (t{-}s)} \Big) \;{\leq}\; \Lambda_{ij} \;{\leq} \;\frac{C_3} {(\lambda(t{-}s))^{3/2}} \exp \Big( {-} C_4 \frac { |x{-}y|^2}{ \lambda (t{-}s)} \Big),
\end{equation}
where $\Lambda_{ij}{=}\Lambda_{ij}(x,t,y,s)$ for every $i,j{=}1,2,3$. The constants, $C_k{=}C_k(T)$, are all positive. In the view of equation (\ref{eq:mag-adj}), we assert that the fundamental solution is unique. Hence ${\Lambda}^{\dagger}(y,s,x,t)$ exists and ${\Lambda}^{\dagger}(y,s,x,t){=}( {\Lambda} (x,t,y,s))'$, where the prime denotes the matrix transpose.

Thus the following bounds hold  ($t{>}0$)
\begin{equation} \label{eq:mag-lq-lp}
 \begin{split}
	\| B(\cdot,t)\|_{q} & \: {\leq} \: C \: (\lambda t)^{-3/2\:(1/p-1/q)} \: \| B_0 \|_{p}, \\
		\| \nabla B(\cdot,t)\|_{q} & \: {\leq} \: C \: (\lambda t)^{-3/2\:(1/p-1/q)-1/2} \: \| B_0 \|_{p},
 \end{split}
\end{equation}
where $1{\leq}p{\leq}q{\leq}{\infty}$ except the case $p{=}1, q{=}\infty$, and $C{=}C(p,q,T,B_0)$. 

As mentioned earlier, the curl of the Lorentz force $J{\times}B$ is a couple and it affects the angular velocity according to the principle of conservation of angular velocity. The first equation in (\ref{eq:curls}) can be written in the following form so that the moment of the Lorentz force is considered as a ``forcing'' term on the right-hand side:
\begin{equation} \label{eq:vort}
	\partial_t \omega_i - \nu \Delta \omega_i +  (u . \nabla )\omega_i - (\omega . \nabla) u_i  = {\mathscr F}\big(B, B_0\big).
\end{equation}
The differential equation on the left is the vorticity equation in fluid dynamics. The global regularity of the solution has been established (Lam 2013). Thus the vorticity for (\ref{eq:vort}) satisfies the integral equation
\begin{equation} \label{eq:pde-vort-sol}
	\omega(x,t) = \int {\Gamma}(x,t,y) \; \omega_0(y) \rd y + \int_0^t \!\! \int {\Gamma}(x,t,y,s) {\mathscr F}\big(B,B_0\big)(y,s) \rd y \rd s,
\end{equation}
where ${\Gamma}$ is the fundamental solution of the vorticity equation.  

Let ($\tilde{B}$, $\tilde{\omega}$, $\tilde{u}$) be a second motion for the MHD equations with the initial data ($\tilde{B}_0$, $\tilde{\omega}_0$, $\tilde{u}_0$). 
Evidently, we have
\begin{equation} \label{eq:mag-sol}
	B(x,t) - \tilde{B}(x,t) = \int \Big( \Lambda(x,t,y) B_0(y) - \tilde{\Lambda}(x,t,y) \tilde{B_0}(y) \Big) \rd y.
\end{equation}
Then it follows that
\begin{equation} \label{eq:vort-sol}
\begin{split}
	\omega(x,t) - \tilde{\omega}(x,t) = \int & \Big( \Gamma(x,t,y)\omega_0(y) - \tilde{\Gamma}(x,t,y)\tilde{\omega}_0(y) \Big) \rd y \\ 
	& + \int_0^t \!\! \int \Big( \Gamma(x,t,y,s) {\mathscr F}\big(B,B_0\big)(y,s)  \\
	 & {\hspace{1.5cm}} - \tilde{\Gamma}(x,t,y,s) {\mathscr F}\big(\tilde{B},\tilde{B}_0\big)(y,s) \Big) \rd y \rd s.
\end{split}
\end{equation}
If the initial velocity data coincide, then $\Lambda{=} \tilde{\Lambda}$. Hence $B{=}\tilde{B}$ if the initial magnetic fields have identical values. It follows that the difference in the vorticity remains zero if the initial vorticity data coincide.

\subsection{Decay in large time}

As $t \rightarrow \infty$, the vorticity and magnetic fields both decay due to the dissipations, as there are no external energy sources. The final decays are well approximated by an equation of diffusion type. By analogous analysis of the long-time decay for the Navier-Stokes dynamics, we can estimate some long-time decays:
\begin{equation} \label{eq:vort-decay}
	{\|\omega(\cdot,t) \|_{L^\infty(\real^3)}}  \:\; {\sim} \:\;  (\nu t)^{-5/4} \; {\| u_0 \|_{L^2(\real^3)}},
\end{equation}
and
\begin{equation} \label{eq:vel-decay}
	 {\| u(\cdot,t) \|_{L^\infty(\real^3)}}  \:\; {\sim} \:\;  (\nu t)^{-3/4} \; {\| u_0 \|_{L^2(\real^3)}}, \;\;\;\;\;\; \mbox{as}\;\;\; t \rightarrow \infty.
\end{equation}
The induction equation in (\ref{eq:mhd}) reduces to ${\partial B}/{\partial t} {-} \lambda \Delta B {=} 0$ which has the well-known bound (\ref{eq:mag-lq-lp}). Thus the decay is given by
\begin{equation} \label{eq:mag-decay}
	{\|B(\cdot,t) \|_{L^\infty(\real^3)}}  \:\; {\sim} \:\;  (\lambda t)^{-3/4} \; {\| B_0 \|_{L^2(\real^3)}}\;\;\;\;\;\; \mbox{as}\;\;\; t \rightarrow \infty.
\end{equation}

\section{Approximate solutions}

It is apparent that {\itshape constructions} of general solutions to the system (\ref{eq:curls}) for arbitrary initial conditions remain a very difficult task. However, there are a number of possibilities where approximations can be made according to the physics involved. Consequently, these approximate solutions can be solved to provide insights into the magneto-hydrodynamic problems. First we derive the equivalent integral equations.

\subsection{System of integral equations}

The first equation in (\ref{eq:curls}) reduces to
\begin{equation} \label{eq:vort-int-t}
	\omega_i(x,t) =  f_i(x,t) + 
	\int_0^{t} \!\! \int \Big(  \big( \omega_i u {-} u_i\omega + (B_i J {-}J_i B)/\rho \big).\nabla_y\Big) Z(x,t,y,s) \: \rd y \rd s.
\end{equation}
The magnetic field $B$ is the induced effect of the current, as shown in the Biot-Savart relation (\ref{eq:bs}). Similarly, the velocity is induced by the vorticity. 
Combining with (\ref{eq:bs-vel}) and (\ref{eq:bs}), the components of $\omega$ can be rewritten as
\begin{equation} \label{eq:vort-t}
 \begin{split}
	\omega_i(x,t) & = f_i(x,t) + \int_0^{t} \!\! \int \!\! \int   \sum_{l=1}^3 G_{il}(x,t,y,s,z,s) \omega_l(z,s)\omega(y,s)  \rd z \rd y \rd s   \\
	& {\hspace{2.5cm}}+ \int_0^{t} \!\!\! \int \!\! \int   \sum_{l=1}^3 F_{il}(x,t,y,s,z,s) J_l(z,s) J(y,s)  \rd z \rd y \rd s,
 \end{split}
\end{equation}
where the scaled coefficient matrix $F = - {\mu_0}G/{\rho}$. Similarly, the second equation in (\ref{eq:curls}) for the magnetic field is transformed into
\begin{equation} \label{eq:mag-int-t}
	B_i(x,t) =  b_i(x,t) + 
	\int_0^{t} \!\! \int \Big( (B_i u - u_i B).\nabla_y \Big)Y(x,t,y,s) \: \rd y \rd s.
\end{equation}
or into its vector form
\begin{equation} \label{eq:mag-t}
	B(x,t) = b(x,t) + \int_0^{t} \!\!\! \int Q(x,t,y,s,\omega) B(y,s)\rd y \rd s,
\end{equation}
where the elements of $Q$ are found from 
\begin{equation*}
	Q_{ij} = \int G'_{ij}(x,t,y,s,z) \omega(z) \rd z.
\end{equation*}
The kernel $G'$ has the elements $(\alpha', \beta', \gamma')$, and they are the elements of $G$ with every $Z$ replaced by $Y$. 
In terms of the current density, equation (\ref{eq:mag-t}) is expressed as
\begin{equation} \label{eq:density-k}
\begin{split}
	J(x,t) & = d(x,t) + \int_0^{t} \!\!\! \int R(x,t,y,s,z,\omega) B(y,s) \rd y \rd s,
\end{split}
\end{equation}
where $d = \nabla {\times} b/\mu_0$, and the kernel $R = \nabla{\times}Q/\mu_0$.
We use the notations, $R_{ij}$, to denote the row matrices of the form
\begin{equation*}
	R_{ij}=(\; \alpha''_{ij} \;\;\; \beta''_{ij} \;\;\; \gamma''_{ij} \;).
\end{equation*}
By the rule of the curl, it is easy to see that, for instance,
\begin{equation*}
	R_{11}=(\; \alpha''_{11} \;\;\; \beta''_{11} \;\;\; \gamma''_{11} \;)=  (\;\partial_{x_2}\alpha'_{31}{-}\partial_{x_3}\alpha'_{21}\;\;\partial_{x_2}\alpha'_{32}{-}\partial_{x_3}\alpha'_{22}\;\;\partial_{x_2}\alpha'_{33}{-}\partial_{x_3}\alpha'_{23}\;),
\end{equation*}
where all the space derivatives operate only on $Y$.

Next we need some knowledge of the integral kernels involved. The derivative of the diffusion kernel $Z$ is bounded by
\begin{equation} \label{eq:dz-bound}
	\big| \partial_{x_i} Z (x,t,y,s) \big| \leq C_1 (\nu t {-} \nu s)^{-\nu_1} |x{-}y|^{-4+2 \nu_1} \exp\Big( {-} \nu_1^* \frac{|x{-}y|^2}{4 \nu (t{-}s)}\Big),
\end{equation}
where $C_1$ is constant, $0{\leq}\nu_1{\leq}2$, and $0{<}\nu_1^*{<}1$. Similarly, it is known that
\begin{equation} \label{eq:d2z-bound}
	\big| \partial^2 Z (x,t,y,s)/\partial_{x_i} \partial_{x_j}\big| \leq C_2 (\nu t {-} \nu s)^{-\nu_2} |x{-}y|^{-5+2 \nu_2} \exp\Big( {-} \nu_2^* \frac{|x{-}y|^2}{4 \nu (t{-}s)}\Big),
\end{equation}
where $C_2$ is constant, $0{\leq}\nu_2{\leq}3/2$, and $0{<}\nu_2^*{<}1$ (see, for example, Friedman 1964; Edel'man 1969). Similar bounds hold for the fundamental solution $Y$.

\subsection{Weak induced electrical field}

If the motion of the conducting fluid is set up in such a manner that the induced electrical field is rather weak or $u{\times}B$ remains zero, then
\begin{equation} \label{eq:curls-bb}
 \begin{split}
	\partial_t \omega_i - \nu \Delta \omega_i & = (\omega . \nabla) u_i  - (u . \nabla )\omega_i  + \Big((B . \nabla) J_i  - (J . \nabla )B_i \Big)/\rho, \\
	\partial_t B_i - \lambda \Delta B_i& = 0.
	\end{split}
\end{equation}
We have already treated this problem in the preceding section. On the basis of expression (\ref{eq:f1}), we assert that the effect of the magnetic field is equivalent to a modification of the initial mollified vorticity data (cf. ({\ref{eq:omega1}})).

\subsection{Small fluid conductivity}

If the conductivity is high, the current or the induced magnetic field is weak according to Ohm's law. The ``forcing'' part in the induction is dominated by the imposed magnetic field (say $\Phi$). Moreover any spatial variations of this imposed field in the Lorentz force can be neglected; the Lorentz force is assumed to be proportional to the velocity. Hence the equations of motion are approximately given by 
\begin{equation} \label{eq:curls-b0}
 \begin{split}
	\partial_t \omega_i - \nu \Delta \omega_i & = (\omega . \nabla) u_i  - (u . \nabla )\omega_i  - (\Phi. \nabla)J  /\rho, \\
	\partial_t B_i - \lambda \Delta B_i & =   - (\Phi. \nabla)u_i.
	\end{split}
\end{equation}
The second equation is reduced to
\begin{equation} \label{eq:mag-b0}
	J(x,t) = d(x,t)+ \int_0^t \!\! \int M(x,t,y,s) \omega(y,s)\rd y \rd s
\end{equation}
where the kernel $M=M(Z,Y,\Phi)$. The vorticity equation can be transformed into the form
\begin{equation*}
\begin{split}
	\omega_i(x,t) = f_i(x,t) + g_i(x,t) + \int_0^t \!\! \int \!\! \int \sum_{i=1}^3 G_{ij}(x,t,&y,s,z,s) \omega_j(z,s) \omega(y,s) \rd z \rd y \rd s \\
	& + \int_0^t \!\! \int D(x,t,y,s) \omega(y,s) \rd y \rd s
\end{split}
\end{equation*}
in view of (\ref{eq:mag-b0}). The additional initial mollified data, $g$, is a function of $\Phi$ and $d$. The linear kernel $D=D(\Phi,Y)$ which couples Ohmic dissipation with the velocity field. In the component-wise form, this integral equation can be represented as
\begin{equation} \label{eq:mag-ie-b0}
	\omega_i(x,t) = \bar{f}_i(x,t) + \int_0^t \!\! \int \!\! \int \sum_{i=1}^3 A_{ij}(x,t,y,s,z,s) \omega_j(z,s) \omega(y,s) \rd z \rd y \rd s,
\end{equation}
where the initial condition $\bar{f}$ is filtered $f+g$ by the resolvent of $D$. Similarly, the new integral kernel $A_{ij}$ denotes the filtered $G_{ij}$. Thus equation (\ref{eq:mag-ie-b0}) is in the identical form to the non-linear integral equation governing the vorticity in the Navier-Stokes dynamics. Its solution has an analogous form to (\ref{eq:vort-series-sol}).

\subsection{Linear dependence on vorticity}

One obvious way to solve our problem is to solve (\ref{eq:mag-t}), so that $B$ is expressed in terms of $\omega$ in the Volterra series:
\begin{equation} \label{eq:mag-resolvent}
	B(x,t) = b(x,t) + \int_0^t \!\! \int \!\! Q^*(x,t,y,s,\omega) b(y,s) \rd y \rd s,
\end{equation}
where $Q^*$ is the resolvent of $Q$. The treatment of the general problem is clearly a prohibitive task. In particular, our non-linearity reduction does not seem to work any more because the resolvent becomes a ``polynomial'' of convolutions in $\omega$. To get some insight into the way the flow and magnetic fields interact with each other, we consider an idealized scenario. Suppose that
\begin{equation} \label{eq:mag-linear}
	B_i(x,t) = b_i(x,t) + \int_0^t \!\! \int \!\! \int \sum_{j=1}^3 G'_{ij}(x,t,y,s,z,s) \omega_j(z,s) b(z,s) \rd z \rd y \rd s.
\end{equation}
The last term is linear in the vorticity. Then the integral related to the current density in (\ref{eq:vort-t}) can be evaluated.  The result is given in the following equation
\begin{equation} \label{eq:vort-linear}
 \begin{split}
	\omega_i(x,t) = f_i(x,t) & + g_i(x,t) + \int_0^{t} \!\! \int \sum_{j=1}^3 S_{ij}(x,t,y,s) \omega_j(y,s) \rd y \rd s \\
	& + \int_0^t \!\! \int \!\! \int \sum_{j=1}^3 H_{ij}(x,t,y,s,z,s) \omega_j(z,s)\omega(y,s)  \rd z \rd y \rd s,
 \end{split}
\end{equation}
where
\begin{equation*}
	g_i(x,t) = \int_0^t \!\! \int \!\! \int \sum_{j=1}^3 F_{ij}(x,t,y,s,z,s) d_j(z,s) d(y,s)  \rd z \rd y \rd s.
\end{equation*}
The kernel $H$ depends on $d^2$ as it is the sum of kernel $G$ and the kernel arising from the product of the linear term in (\ref{eq:mag-linear}) with itself. The kernel $S$ comes from the product of the linear term and the initial current density. Let $S^*$ be the resolvent of $S$.
Thus this non-linear integral equation can be reduced to
\begin{equation} \label{eq:vort-linear-res}
	\omega_i(x,t) = h_i(x,t) 
	+ \int_0^t \!\! \int \!\! \int \sum_{j=1}^3 W_{ij}(x,t,y,s,z,s) \omega_j(z,s)\omega(y,s)  \rd z \rd y \rd s,
\end{equation}
where $W$ denotes the convoluted kernel of $H$ and $S^*$. The initial condition $h$ is the filtered data by $S^*$: 
\begin{equation*}
	h = \int_0^t \!\! \int S^*(x,t,y,s) ( f + g )(y,s) \rd y \rd s.
\end{equation*}
The effects due to the linear term in (\ref{eq:mag-linear}) are included in the kernels $S$ and $H$. Equation (\ref{eq:vort-linear-res}) is essentially in the identical form compared to the vorticity equation in the Navier-Stokes dynamics. Thus we assert that the solution has an analogous series expansion
\begin{equation} \label{eq:linear-vort-series-sol}
	\omega(x,t) = \hat{\gamma}(x,t) + \sum_{m{\geq}2}\: {S_m} \: \hat{V}[\hat{\gamma}(x,t)]^m.
\end{equation}
Hence the presence of a magnetic field affects the core of every eddy in the above series and the magnetic field in turn modifies every space-time integral convolution; all the eddies are magnetized, as expected.

The series defines an intricate spatio-temporal vorticity population. It is known as turbulence. The initiation of the turbulence has nothing to do with instability or the presence of a singularity in space or in time. (Similarly, the solutions given in (\ref{eq:vort-series-sol}) and (\ref{eq:mag-ie-approx-sol}) consist in a structure of magneto-hydrodynamic turbulence which is not a result of instability.) A series solution for the MHD equations, such as (\ref{eq:linear-vort-series-sol}), is a direct consequence of the non-linearity in the equations, while the principal effect of the magnetic field is embedded in the core of the vorticity eddies. In general, every eddy has its own distinct strength and scale. The presence of a magnetic field adds an extra degree of complexity to this intricate turbulent field, as we may interpret the effect of the magnetic field as an additional torque on the vorticity (cf. (\ref{eq:vort})). Suppose that the size of the initial data is large, then the vorticity and magnetic fields will be observed as a laminar dynamics, in a short time interval from the start of the motion. As the time goes by, the laminar flow field can no longer sustain and it becomes a turbulent flow via the process of the laminar-turbulent transition in an analogous manner to the Navier-Stokes flow field. 

\subsection{Presence of a strong Lorentz force}

If a magnetic field $B$ is large-scale and strong, its effect on the energy transfer process in small scales was discussed by Kraichnan (1965). In the present approximation, if the Lorentz force is much stronger than the inertia force, the MHD equations (\ref{eq:curls}) reduce to
\begin{equation} \label{eq:curls-lorentz}
 \begin{split}
	\partial_t B_i - \lambda \Delta B_i & =   (B. \nabla) u_i - (u . \nabla) B_i,\\
	\partial_t \omega_i - \nu \Delta \omega_i & =  \Big((B . \nabla) J_i  - (J . \nabla )B_i \Big)/\rho.
	\end{split}
\end{equation}
Either differential equation can be transformed into an integral equation, namely
\begin{equation} \label{eq:mag-ie-lorentz}
	J_i(x,t) = d_i(x,t) + \int_0^{t} \!\! \int \!\! \int \sum_{j=1}^3 V_{ij}(x,t,y,s,z,s) \omega_j(z,s) J(y,s) \rd z \rd y \rd s,
\end{equation}
where $V=RF$, and
\begin{equation} \label{eq:vort-ie-lorentz}
	\omega_i(x,t) = f_i(x,t) + \int_0^t \!\! \int \!\! \int \sum_{l=1}^3 F_{il}(x,t,y,s,z,s) J_l(z,s) J(y,s) \rd z \rd y \rd s.
\end{equation}
Since the vorticity field is expected to be weak, we can eliminate $\omega$ from these two integral equations. The result is a non-linear integral equation in the current density. In fact, it is a cubic power in the current density. The explicit equation reads
\begin{equation} \label{eq:mag-lorentz}
 \begin{split}
 J_i(x,t) = \theta_i(x,t) + \int_0^t \!\! \int_r^s \!\! \int \!\! \int \!\! \int \Big( \sum_{k=1}^3 & E_{ik}(x,t,y,s,z,r,z',r) \\
 & J_k(z',r) J(z,r) J(y,s) \Big) \rd z' \rd z \rd y \rd r \rd s,
	\end{split}
\end{equation}
where $\theta$ is the filtered initial data $d$ by the resolvent of $Vf$, 
\begin{equation*}
	\theta(x,t) = d(x,t) + \int_0^t \!\! \int {\mathscr R}(Vf)(x,t,y,s) d(y,s)  \rd y \rd s,
\end{equation*}
and the kernel $E$ is the convoluted kernel of $F$ and the resolvent ${\mathscr R}(Vf)$.
The construction of the solution for integral equation (\ref{eq:mag-lorentz}) is much more involved. Nevertheless, we find that the solution is given by
\begin{equation} \label{eq:mag-series}
 \begin{split}
	J(x,t) & {=}\:  \theta(x,t) \: + \sum_{m=1}\: {T_m} \: M[\theta(x,t)]^{2m+1}  \\
	\quad & \\
	\quad & {=} \: \theta + 2 \: M[\theta]^3 + 14 \: M[\theta]^5 + 130 \: M[\theta]^7 + 1382 \: M[\theta]^9 + 15906 \: M[\theta]^{11} \\ 
	 \quad & \\
	\quad & {\hspace{5mm}} +  192894 \: M[\theta]^{13} + 2427522 \: M[\theta]^{15} + 31405430 \: M[\theta]^{17} +   \cdots \;,
 \end{split}
\end{equation}
where $M[\theta]^{2m+1}$ denotes the integro-powers terms of degree $2m{+}1$ in the initial data $\theta$. In \ref{app:a}, we list the leading $31$ coefficients. In brief, the series solution shows that, as a result of the cubic non-linearity, the integro-power terms or small scale motions in the magnetic field are more abundant than those in the vorticity field (cf. (\ref{eq:vort-series-sol})).

In the induction equation in the presence of a strong magnetic field (cf. (\ref{eq:curls-lorentz})), there are cancellations in the non-linear terms $(B.\nabla)u$ and $(u.\nabla)B$ when the equation is written as an integral equation. The first term is only responsible for the stretching of magnetic field lines. In fact, the stretching processes do not directly instigate the scale hierarchy of the motion. The resulting motion is a typical turbulent motion with the presence of multitudinous small scales, which are responsible for enhanced diffusivity. 

\section{Some Remarks on MHD Turbulence} 

Multiplying the second equation in (\ref{eq:curls-total}) by $\Pi$ and integrate the result over space and time, we obtain an identity which expresses the law of conservation of energy:
\begin{equation} \label{eq:energy}
	\frac{1}{2}\int \Pi^2 \rd x + \lambda \int_0^t \!\! \int (\nabla \Pi)^2 \rd x \rd s = \frac{1}{2}\int \Pi^2_0 \rd x.
\end{equation}
It is easy to assert that the enstrophy is also a conserved quantity of the motion as
\begin{equation} \label{eq:enstrophy}
	\int \Omega^2 \rd x + 2 \nu \int_0^t \!\! \int (\nabla \Omega)^2 \rd x \rd s = \int \Omega^2_0 \rd x.
\end{equation}
From these conservation laws, we deduce some additional flow quantities which elucidate the nature of turbulence in magnetohydrodynamics. Equation (\ref{eq:energy}) has an alternative form 
\begin{equation} \label{eq:mag-enstrophy}
	\frac{\rd }{\rd t} \int \Pi^2 \rd x = - 2 \lambda \int (\nabla \Pi)^2 \rd x \leq 0,
\end{equation}
where the last equality sign holds only when the magnetic diffusivity $\lambda$ vanishes. A similar bound holds for the vorticity
\begin{equation} \label{eq:vort-enstrophy}
	\frac{\rd }{\rd t} \int \Omega^2 \rd x = - 2 \nu \int (\nabla \Omega)^2 \rd x \leq 0.
\end{equation}
It is clear that we can calculate similar bounds for the derivatives such as $\partial_x^{\alpha} \partial_t^{\beta} \Pi$. In particular, we are interested in two motions with fluctuations in their initial data. We readily assert that
\begin{equation} \label{eq:delta-mag}
	\int (\Pi - \tilde{\Pi})^2 (x,t) \rd x \:{\leq}\: \int (\Pi_0 - \tilde{\Pi}_0)^2(x) \rd x,
\end{equation}
and 
\begin{equation} \label{eq:d-delta-mag}
\int \big( \nabla \Pi - \nabla \tilde{\Pi} \big) ^2 (x,t) \rd x \;{\leq}\; \int \big( \nabla \Pi_0 - \nabla \tilde{\Pi}_0 \big)^2(x) \rd x.
\end{equation}
Both the inequalities are independent of the diffusivity. In short, the fluctuations in the initial conditions are not amplified over the flow evolution. Therefore, the development of the MHD dynamics is {\em insensitive} to small variations in the initial conditions. This is a generalization of the vorticity dynamics in the Navier-Stokes system (Lam 2013). Thus, the magnetohydrodynamics equations are deterministic in nature and they do not embody the characters of a typical chaos system.

Lam (2013) has made an attempt to explain the perceived macroscopic randomness on the basis of the Boltzmann equation for Maxwellian molecules with cut-off. In the MHD equations, both the viscous and Ohmic dissipations are evident, and they are the only irreversible processes during the dynamic evolution. There must be some mutual interaction between the two dissipative processes which is difficult to quantify. By analogy, we deduce that the apparent randomness in MHD has its microscopic origin; the Ohmic dissipation on the small-scale magnetic-vorticity eddies must be irregular. The random fluctuations on the dissipative scales are magnified and passed on, by the Biot-Savart action, to the whole flow field, so long as the dissipations are intensive. 

\section{Conclusion}

It is shown in the present paper that the magneto-hydrodynamic equations in $\real^3$ are globally well-posed for any finite-energy initial data which are smooth and localized. The mean flow quantities are well described by the equations. In the MHD system of a weak Lorentz force, the dominant effect is due to the influence of the fluid dynamics; turbulence is an inherent feature of the dynamics if the size of the initial data are considerable. On the other hand, if a strong Lorentz force is present, the approximate dynamic equations contain a cubic non-linearity in the current density. It is shown that the non-linear integral equation can also be solved in a series expansion. The key ideas in the present paper are a generalization of the turbulent flows governed by the Navier-Stokes equations of motion. However, the interactions between the velocity and magnetic fields are exceedingly complex.


\appendix{List of the coefficients $T_m$} \label{app:a}

Selected coefficients in the approximation for strong magnetic field are listed below\footnote{To appear as {\ttfamily A235347} in {\ttfamily www.oeis.org}}. 

\begin{center}
\begin{tabular}{ccl} \hline \hline
$\;\;\;m\;\;\;$ & $\;\;\;2m{+}1\;\;\;$   & $T_m$ in Equation (\ref{eq:mag-series}) \\ \hline \hline
1 & 3   & 2 \\
2 & 5   & 14 \\
3 & 7   & 130 \\
4 & 9   & 1382 \\
5 & 11  & 15906 \\
6 & 13  & 192894 \\
7 & 15  & 2427522 \\
8 & 17  & 31405430 \\
9 & 19  & 415086658 \\
10& 21  & 5580629870 \\
11& 23  & 76080887042 \\
12& 25  & 1049295082630 \\
13& 27  & 14613980359010 \\
14& 29  & 205246677882078 \\
15& 31  & 2903566870820610 \\
16& 33  & 41337029956899222 \\
17& 35  & 591796707042765954 \\
18& 37  & 8514525059135909070 \\
19& 39  & 123048063153362454402 \\ 
20& 41  & 1785343913603396041638 \\
21& 43  & 25997755243402345012386 \\
22& 45  & 379816229907918156775998 \\
23& 47  & 5565580921491549002988930 \\
24& 49  & 81778291342796926262452662 \\
25& 51  & 1204648803858660621476918466 \\
26& 53  & 17786610464868727152485460014 \\
27& 55  & 263184989591618354500003475458 \\
28& 57  & 3902088960176965421506004890310 \\
29& 59  & 57961718392401483193930375000034 \\
30& 61  & 862460700267404862132628470059422 \\
31& 63  & 12854125699306884541089041186674178\;\; \\ \hline \hline
\end{tabular}
\end{center}

\vspace{4mm}

\noindent 
13 June 2023.

\noindent 
\texttt{f.lam11@yahoo.com}

\label{lastpage}
\end{document}